\documentclass[a4paper,onecolumn,accepted=2023-06-07]{quantumarticle}
\pdfoutput=1

\usepackage[usenames,dvipsnames]{xcolor}
\usepackage{amsmath,amssymb,amsfonts,amsthm}
\usepackage{graphicx}
\usepackage{tabularx}

\usepackage[T1]{fontenc}
\usepackage[utf8]{inputenc}

\usepackage{csquotes}
\usepackage{nicefrac}
\usepackage[textsize=tiny]{todonotes}
\usepackage{booktabs}
\usepackage[hyphens]{url}
\usepackage{hyperref}
\usepackage{tikz}
\usetikzlibrary{arrows,positioning} 
\tikzset{
    >=stealth',
    punkt/.style={
           rectangle,
           rounded corners,
           draw=black, very thick,
           text width=6.5em,
           minimum height=2em,
           minimum width=2.5cm,
           text centered},
    pil/.style={
           ->,
           thick,
           shorten <=2pt,
           shorten >=2pt,}
}
\usetikzlibrary{shapes.geometric, positioning, quantikz}

\newcommand{\dashedLegend}{\raisebox{2pt}{\tikz{\draw[dashed](0,0) -- (5mm,0);}}}
\newcommand{\solidLegend}{\raisebox{2pt}{\tikz{\draw(0,0) -- (5mm,0);}}}

\makeatletter
\let\MYcaption\@makecaption
\makeatother
\usepackage[font=footnotesize]{subcaption}
\makeatletter
\let\@makecaption\MYcaption
\makeatother

\usepackage{pifont}

\newtheorem{example}{Example}

\usepackage[style=ieee, maxnames=6, minnames=1, date=year, url=false, isbn=false, backend=biber, sortcites]{biblatex}

\DeclareFieldFormat
  [misc]
  {title}{\mkbibquote{#1}}

\AtEveryBibitem{%
  \clearname{editor}%
  \clearfield{series}%
  \clearfield{isbn}%
    \clearfield{volume}%
  \clearfield{number}%
  \clearfield{pages}%
  \clearfield{location}%
  \clearfield{pubstate}%
  \clearfield{eprintclass}%
  \clearfield{url}%
  \clearfield{urldate}%
}

\usepackage{footnote}
\usepackage{flushend}
\usepackage{yquant}

\begin{document}

\title{MQT Bench: Benchmarking Software and Design Automation Tools for Quantum Computing}

\author{Nils Quetschlich}
\affiliation{Chair for Design Automation, Technical University of Munich, Germany}
\email{nils.quetschlich@tum.de}
\author{Lukas Burgholzer}
\affiliation{Chair for Design Automation, Technical University of Munich, Germany}
\email{lukas.burgholzer@tum.de}
\author{Robert Wille}
\affiliation{Chair for Design Automation, Technical University of Munich, Germany}
\affiliation{Software Competence Center Hagenberg GmbH (SCCH), Austria}
\email{robert.wille@tum.de}

\maketitle

\begin{abstract}
Quantum software tools for a wide variety of design tasks on and across different levels of abstraction are crucial in order to eventually realize useful quantum applications. 
This requires practical and relevant benchmarks for new software tools to be empirically evaluated and compared to the current \mbox{state of the art}. 
Although benchmarks for specific design tasks are commonly available, the demand for an overarching \mbox{cross-level} benchmark suite has not yet been fully met and there is no mutual consolidation in how quantum software tools are evaluated thus far. 
In this work, we propose the \emph{MQT Bench} benchmark suite (as part of the \emph{Munich Quantum Toolkit}, MQT) based on four core traits:
(1)~cross-level support for different abstraction levels, 
(2)~accessibility via an \mbox{easy-to-use} web interface (\url{https://www.cda.cit.tum.de/mqtbench}) and a Python package, 
(3)~provision of a broad selection of benchmarks to facilitate generalizability, as well as 
(4)~extendability to future algorithms, \mbox{gate-sets}, and hardware architectures.
By comprising more than $70{\small,}000$ benchmark circuits ranging from $2$ to $130$ qubits on four abstraction levels, \mbox{MQT Bench} presents a first step towards benchmarking different abstraction levels with a single benchmark suite to increase comparability, reproducibility, and transparency. 
\end{abstract}

\section{Introduction}
Quantum computing has gained a lot of attention due to its promising applications and the advances in quantum computing hardware. But still, designing quantum algorithms often means to manually implement quantum circuits on the gate-level---similar to assembly language in classical computing. 
Hence, there is an urgent need to enhance the workflow of designing and testing quantum algorithms. 
Without software tools that aid in the design of quantum algorithms, the underlying hardware may not be efficiently utilized. 
Thus, researchers and developers have already proposed tools for various design tasks, such as quantum circuit simulation~\mbox{\cite{sim_1, sim_2, hillmichApproximatingDecisionDiagrams2022, hillmichConcurrencyDDbasedQuantum2020, burgholzerHybridSchrodingerFeynmanSimulation2021, burgholzerSimulationPathsQuantum2022, kissingerClassicalSimulationQuantum2022, brennanTensorNetworkCircuit2021, vincentJetFastQuantum2021, imGraphPartitioningApproach2023, lykovTensorNetworkQuantum2020, deraedtMassivelyParallelQuantum2019, bravyiImprovedClassicalSimulation2016}}, compilation~\mbox{\cite{compilation_1, compilation_2, compilation_3, quetschlich2023prediction, quetschlich2023compileroptimization, pehamDepthoptimalSynthesisClifford2023,
burgholzerLimitingSearchSpace2022,
willeMappingQuantumCircuits2019,
hillmichExploitingQuantumTeleportation2021,
zulehnerEfficientMethodologyMapping2019,
zulehnerCompilingSUQuantum2019,
shaikOptimalLayoutSynthesis2023,
liuTacklingQubitMapping2023,
willeMQTQMAPEfficient2023,
zhangTimeoptimalQubitMapping2021,
muraliNoiseadaptiveCompilerMappings2019,
cowtanQubitRoutingProblem2019,
sivarajahKetRetargetableCompiler2021,
tanOptimalLayoutSynthesis2020,
liTacklingQubitMapping2019,
revlib_example_2}}, or verification~\mbox{\cite{verification_2, verification_3, verification_4, verification_5, pehamEquivalenceCheckingQuantum2022,
pehamEquivalenceCheckingParameterized2023,
burgholzerAdvancedEquivalenceChecking2021,
burgholzerRandomStimuliGeneration2021,
burgholzerHandlingNonunitariesQuantum2022,
chun-yuAccurateBDDbasedUnitary2022,
taoGiallarPushbuttonVerification2022,
willeVerificationQuantumCircuits2022,
wangXQDDbasedVerificationMethod2008,
hongApproximateEquivalenceChecking2021,
chenAutomatabasedFrameworkVerification2023,
liuRobustApproachDetecting2023}}. 
This already led to comprehensive quantum circuit design flows realized through toolkits such as IBM's Qiskit~\cite{qiskit}, Google's Cirq~\cite{cirq}, or Rigetti's Forest~\cite{rigetti}---in addition to numerous tools and methods developed by other researchers and engineers in the field.

Generally, these software tools operate on and across different levels of abstraction and all tackle computationally hard problems. As a result, they have to compromise between resource demands and result quality. 
Usually, when a new software tool is proposed, its performance is empirically evaluated. 
The question how to evaluate the performance of a software tool is very challenging and not yet fully answered---especially for tools solving NP-complete problems. 
The most common approach is to run certain problem instances, \mbox{so-called} benchmarks, and compare the performance of the newly proposed method against \mbox{state-of-the-art} methods regarding a certain characteristic, e.g., \mbox{run-time} or solution quality.

Currently, this benchmarking is conducted using a wide variety of different benchmark suites---each with a specific focus. While some benchmarks suites, e.g., \cite{lubinski2021applicationoriented, supermarq}, provide benchmarks on higher abstraction levels, 
other benchmark suites focus on lower abstraction levels, e.g., \mbox{\cite{li2022qasmbench, revlib}}. Although the intended target levels are properly covered, the demand for a \mbox{cross-level} benchmark suite is not fully met yet. 
As another consequence, there is no mutual consolidation which benchmarks to use for empirical evaluations yet---leading to lower comparability, reproducibility, and transparency.

In this paper, we propose \emph{MQT Bench}---the benchmark suite from the \emph{Munich Quantum Toolkit}~(MQT) which explicitly aims to address those drawbacks.
It aims at providing a first step towards benchmarking the whole quantum software stack with a single benchmark suite by offering the same benchmark algorithms on different levels of abstraction. To realize such a benchmark suite, several challenges have to be tackled:
\begin{itemize}
\item Distinct requirements on different levels: Benchmarks have to fulfill certain requirements depending on the abstraction level, e.g., only gates of a device's native \mbox{gate-set} are allowed.
\item Accessibility: To foster adoption of the benchmark suite, it needs to be as easy to use as possible.
\item Generalizability: Providing a comprehensive set of benchmark algorithms to cover as many use cases as possible.
\item Extendability: Future algorithms, \mbox{gate-sets}, and architectures should be easy to integrate.
\end{itemize}
MQT Bench has been developed with these challenges in mind and is based on four core traits:
\begin{enumerate}
\item Cross-level benchmarking: Provision of benchmarks on four abstraction levels.
\item Accessibility: To simplify the usage of MQT Bench, we provide a web interface (\url{https://www.cda.cit.tum.de/mqtbench}) and a Python package in addition to the \mbox{open-source} repository on GitHub (\url{https://github.com/cda-tum/mqt-bench}).
\item Algorithm selection: Broad selection of different algorithms ranging from building blocks, i.e., Quantum Fourier Transform~(QFT), to applications, i.e., Grover's algorithm.
\item Extendability: Algorithms, \mbox{gate-sets}, and hardware architectures are easily extendable and integrable into MQT Bench. 
\end{enumerate}
By this, \mbox{MQT Bench} aims to improve the comparability, reproducibility, and transparency of empirical evaluations for the whole quantum software stack.

The rest of this work is structured as follows: In \autoref{sec:background}, we review the necessary basics of the quantum computing compilation flow and the respective software stack to keep this work self-contained. Afterwards,
\autoref{sec:benchmarking} reviews the \mbox{state of the art} on how this quantum software stack is currently evaluated and benchmarked---motivating the benchmark suite proposed in this paper. Based on that, \autoref{sec:detail_section} introduces the resulting benchmark suite and its core traits, before the generated benchmarks are evaluated in~\autoref{sec:eval}.
\autoref{sec:conclusion} concludes this work.

\section{Background}
\label{sec:background}
To keep this paper \mbox{self-contained}, this section gives a brief overview of the quantum circuit compilation flow with its different abstraction levels and the respective quantum software stack. The levels mentioned in this section are inspired by the structure proposed by the openQASM 3.0 specification~\cite{cross_openqasm_2021} and can be found similarly in many other compilation flows as well.

\subsection{Quantum Circuit Compilation Flow}
\label{sec:compflow}
Similar to the classical domain, executing a conceptual quantum algorithm on an actual device requires compiling it to a representation that adheres to all constraints imposed by the hardware. Usually, quantum algorithms are initially developed and tested on a hardware-agnostic level. Generally, this is described as a quantum circuit that consists of \mbox{high-level} building blocks without any restrictions to a certain \mbox{gate-set} or hardware architecture and is defined as the \emph{algorithmic level}. 

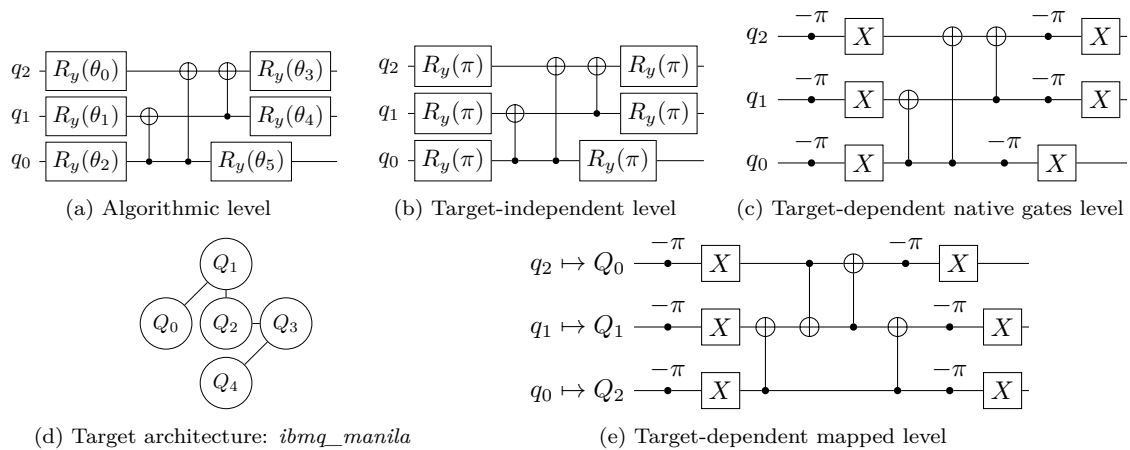
\begin{figure}     
	\begin{subfigure}{0.99\textwidth}     
     \begin{subfigure}{0.3\textwidth}     
		     \resizebox{1.0\linewidth}{!}{
				\begin{tikzpicture}
				  \begin{yquant}
				    	qubit {${q_2}$} q;    	
						qubit {${q_1}$} q[+1];
						qubit {${q_0}$} q[+1];
				    	
				    	box {$R_y(\theta_0)$} q[0];
				    	box {$R_y(\theta_1)$} q[1];
				    	box {$R_y(\theta_2)$} q[2];
				    	
				    	cnot q[1] | q[2];
				    	cnot q[0] | q[2];
				    	cnot q[0] | q[1];
				    	
				    	box {$R_y(\theta_3)$} q[0];
				    	box {$R_y(\theta_4)$} q[1];
				    	box {$R_y(\theta_5)$} q[2];
				  \end{yquant}
				\end{tikzpicture}}
		\caption{Algorithmic level}
         \label{fig:sub_1}
		 \end{subfigure}
     \hfill
\begin{subfigure}{0.3\textwidth}
     \resizebox{1.0\linewidth}{!}{
		\begin{tikzpicture}
			  \begin{yquant}
			    	qubit {${q_2}$} q;    	
					qubit {${q_1}$} q[+1];
					qubit {${q_0}$} q[+1];
			    	
			    	box {$R_y(\pi)$} q[0];
			    	box {$R_y(\pi)$} q[1];
			    	box {$R_y(\pi)$} q[2];
			    	
			    	cnot q[1] | q[2];
			    	cnot q[0] | q[2];
			    	cnot q[0] | q[1];
			    	
			    	box {$R_y(\pi)$} q[0];
			    	box {$R_y(\pi)$} q[1];
			    	box {$R_y(\pi)$} q[2];
			
			  \end{yquant}
		\end{tikzpicture}}
         \caption{Target-independent level}
         \label{fig:sub_2}
     \end{subfigure}
     \hfill
          \begin{subfigure}[b]{0.35\textwidth}
     \resizebox{1.0\linewidth}{!}{
		\begin{tikzpicture}
  		\begin{yquant}
    	qubit {${q_2}$} q;    	
		qubit {${q_1}$} q[+1];
		qubit {${q_0}$} q[+1];
    	
		phase {$-\pi$} q[0];
		phase {$-\pi$} q[1];
		phase {$-\pi$} q[2];
		
		x q[0];
		x q[1];
		x q[2];    	    	
    	   	
    	cnot q[1] | q[2];
    	cnot q[0] | q[2];
    	cnot q[0] | q[1];
    	
		phase {$-\pi$} q[0];
		phase {$-\pi$} q[1];
		phase {$-\pi$} q[2];
		
		x q[0];
		x q[1];
		x q[2];
  \end{yquant}
\end{tikzpicture}}
         \caption{Target-dependent native gates level}
         \label{fig:sub_3}
     \end{subfigure}     
     \end{subfigure}
		     
\centering
	\begin{subfigure}[b]{1.0\textwidth}     
	\centering
          		 \begin{subfigure}[b]{0.35\textwidth}
					\centering
			     \resizebox{0.45\linewidth}{!}{
					   \begin{tikzpicture}
						    \node (p1) at ( 0, 1) [circle, draw]{$Q_0$}; 
						    \node (p2) at ( 1, 0) [circle, draw]{$Q_4$};
						    \node (p3) at ( 1,1) [circle, draw]{$Q_2$};
						    \node (p4) at ( 1,2) [circle, draw]{$Q_1$};
						    \node (p5) at ( 2,1) [circle, draw]{$Q_3$};

						    \begin{scope}[every path/.style={-}]
						       \draw (p1) -- (p4);
						       \draw (p3) -- (p4); 
						       \draw (p3) -- (p5);
						       \draw (p2) -- (p5);
						    \end{scope}  
						\end{tikzpicture}}
			         \caption{Target architecture: \emph{ibmq\_manila}}
			         \label{fig:ibm_arch}
		     \end{subfigure}
     \begin{subfigure}[b]{0.6\textwidth}
   \centering
     \resizebox{0.75\linewidth}{!}{
		\begin{tikzpicture}
  \begin{yquant}
    	qubit {${q_2} \mapsto Q_0$} q;    	
		qubit {${q_1} \mapsto Q_1$} q[+1];
		qubit {${q_0} \mapsto Q_2$} q[+1];
    	
		phase {$-\pi$} q[0];
		phase {$-\pi$} q[1];
		phase {$-\pi$} q[2];
		
		x q[0];
		x q[1];
		x q[2];    	    	
    	   	
    	cnot q[1] | q[2];
    	cnot q[1] | q[0];
    	cnot q[0] | q[1];
    	cnot q[1] | q[2];
    	
		phase {$-\pi$} q[0];
		phase {$-\pi$} q[1];
		phase {$-\pi$} q[2];
		
		x q[0];
		x q[1];
		x q[2];

  \end{yquant}
\end{tikzpicture}}
         \caption{Target-dependent mapped level}
         \label{fig:sub_4}
     \end{subfigure}
     \end{subfigure}
        \caption{Compilation of a Variational Quantum Algorithm.}
\end{figure}

\begin{example}
\emph{Variational Quantum Algorithms} (VQAs) are an emerging class of quantum algorithms with a wide range of applications. A respective circuit is depicted in \autoref{fig:sub_1} and shows an example of an ansatz function frequently used for \emph{Variational Quantum Eigensolvers} (VQEs), a subclass of VQAs. On this abstraction level, the circuit is parameterized by the angles $\theta_i$ of the six \mbox{single-qubit} gates.
\label{ex:1}
\end{example}

The first step in realizing a conceptual quantum algorithm on the algorithmic level for a particular problem is to synthesize the \mbox{high-level} building blocks and optimize the resulting representation independently of the actual target architecture. In analogy to a classical compiler, this involves tasks such as constant propagation and folding, gate modifier evaluation, synthesis, loop unrolling, and gate simplification. The resulting representation is defined as the \mbox{\emph{target-independent level}}.

\begin{example}
VQAs are hybrid quantum-classical algorithms, where the parameters of the quantum ansatz are iteratively updated by a classical optimizer analogous to conventional \mbox{gradient-based} optimization.
Consider again the circuit from \autoref{fig:sub_1}. Assuming that these parameters have been determined, e.g., $\theta_i =\pi$ for $i=0, ..., 5$, they are now propagated and the resulting quantum circuit is shown in \autoref{fig:sub_2}.
\label{ex:2}
\end{example}

Today's quantum devices impose several constraints on the circuits that may be executed on them. 
Thus, a \mbox{target-dependent} compilation phase is necessary. 
For one, devices only provide a particular set of native gates---typically consisting of an entangling gate (like the $CX$ gate) and some family of \mbox{single-qubit} gates. 
Therefore, the gates of a circuit need to be translated to this native \mbox{gate-set}, which is typically followed by an optimization pass that aims to reduce the introduced overhead. 
This representation level is defined as the \mbox{\emph{target-dependent native gates level}}.

\begin{example}
Different quantum computer realizations support different native \mbox{gate-sets}. 
In our example, we consider the \emph{ibmq\_manila} device as the target device that natively supports $I$, $X$, $\sqrt{X}$, $R_z$, and $CX$ gates. 
Consequently, the $R_y$ gates in \autoref{fig:sub_2} have to be converted using only these native gates. 
In this case, they are substituted by a sequence of $R_z$ (denoted as $\bullet$ with a phase of $-\pi$) and $X$ gates as shown in \autoref{fig:sub_3}.
\label{ex:3}
\end{example}

In addition to the limited \mbox{gate-set}, today's devices (at least those based on superconducting qubits) only feature limited connectivity between their qubits. 
Consequently, the circuit's logical qubits need to be \emph{mapped} to the targeted device's physical qubits, so that multi-qubit gates are only applied to qubits directly connected on the device. 
Since such a mapping can rarely be determined in a static fashion, i.e., globally for the whole circuit, the mapping has to change dynamically throughout the circuit, which is frequently referred to as mapping or routing.
Again, this is followed by a round of optimization in order to minimize the overhead caused by the compilation. This representation level is defined as the \mbox{\emph{target-dependent mapped level}}.

\begin{example}
Consider again the scenario from \autoref{ex:3}. The architecture of the \emph{ibmq\_manila} device is shown in \autoref{fig:ibm_arch} and defines between which qubits a two-qubit operation can be performed.
Since the circuit shown in \autoref{fig:sub_3} contains $CX$ gates operating between all combinations of qubits, there is no mapping directly matching the target architecture's layout. As a consequence, a \mbox{non-trivial} mapping followed by a round of optimization leads to the resulting circuit shown in \autoref{fig:sub_4}. This is also the reason for the different sequence of $CX$ gates compared to the previous example.
\end{example}

At this point, the circuit is ready to be sent to the quantum computer's backend for execution. From there, the individual gates are scheduled, linked to a particular calibration, and the resulting circuit is eventually passed to the target machine code generator. The resulting binaries are then forwarded to the execution engine to orchestrate the quantum computation on the actual device.

\subsection{Quantum Software Stack}
Quantum software is needed on all levels reviewed above in order to aid designers in eventually realizing useful quantum applications. In the following, we review three of the core design tasks for software tools.

Today, quantum computers are a scarce resource with limited availability and capability. Until more devices become available that are larger and less prone to errors, classical means to simulate quantum algorithm circuits (also known as \emph{quantum circuit simulators}) are essential for fostering the development of quantum applications. 
Additionally, and in contrast to actual quantum computing devices, quantum circuit simulators allow for detailed insights into the quantum states throughout the circuit execution since the state's amplitudes are explicitly tracked during the simulation. 
On actual hardware, information can only be extracted in the form of measurements from the final quantum state, which each yields a classical bit string according to the distribution described by the state's amplitudes.
Such simulators can be used on the highest possible level of abstraction, since they can be designed in a way that does not require restrictions on the circuit's \mbox{gate-set} or the qubits' connectivity, e.g., as proposed in \cite{sim_1, sim_2, hillmichApproximatingDecisionDiagrams2022, hillmichConcurrencyDDbasedQuantum2020, burgholzerHybridSchrodingerFeynmanSimulation2021, burgholzerSimulationPathsQuantum2022, kissingerClassicalSimulationQuantum2022, brennanTensorNetworkCircuit2021, vincentJetFastQuantum2021, imGraphPartitioningApproach2023, lykovTensorNetworkQuantum2020, deraedtMassivelyParallelQuantum2019, bravyiImprovedClassicalSimulation2016}.
Furthermore, they might also be used on the lowest possible level in order to perform \mbox{noise-aware} simulations to estimate how a circuit is likely to perform on an actual device, e.g., as proposed in \cite{err_sim_1, err_sim_2, err_sim_3, grurlNoiseawareQuantumCircuit2023}.
Either way, on both abstraction levels, quantum circuit simulation is \mbox{non-trivial} since corresponding representations of states and operations, in general, require exponential space or runtime---requiring powerful software tools.

Due to the immense complexity of the tasks involved in quantum circuit compilation (as it has been illustrated in \autoref{sec:compflow}), e.g., mapping being \mbox{NP-complete~\cite{siraichi2018qubit}}, manually conducting compilation often is not an option. 
Consequently, efficient software tools are needed across all levels of abstraction of the aforementioned flow in order to realize a conceptual algorithm on an actual device. 
Thus, various compilation tools have been proposed, e.g., \cite{compilation_1, compilation_2, compilation_3, quetschlich2023prediction, quetschlich2023compileroptimization,pehamDepthoptimalSynthesisClifford2023,
burgholzerLimitingSearchSpace2022,
willeMappingQuantumCircuits2019,
hillmichExploitingQuantumTeleportation2021,
zulehnerEfficientMethodologyMapping2019,
zulehnerCompilingSUQuantum2019,
shaikOptimalLayoutSynthesis2023,
liuTacklingQubitMapping2023,
willeMQTQMAPEfficient2023,
zhangTimeoptimalQubitMapping2021,
muraliNoiseadaptiveCompilerMappings2019,
cowtanQubitRoutingProblem2019,
sivarajahKetRetargetableCompiler2021,
tanOptimalLayoutSynthesis2020,
liTacklingQubitMapping2019,
revlib_example_2}.

During compilation, an algorithm's description is considerably altered and transformed. 
Naturally, it is of utmost importance to ensure that the originally intended functionality is preserved through all levels of abstraction. 
This procedure is called verification. 
While the underlying principle of comparing the transformations represented by different quantum circuits is conceptually simple, the exponential size of the underlying matrices makes this problem challenging---it has been shown to be QMA-complete~\cite{janzingNonidentityCheckQMAcomplete2005}. 
Due to the increasing complexity of today's compilation flows, tools for verifying their results become increasingly important. Examples of verification tools have been proposed in \cite{verification_2, verification_3, verification_4, verification_5, pehamEquivalenceCheckingQuantum2022,
pehamEquivalenceCheckingParameterized2023,
burgholzerAdvancedEquivalenceChecking2021,
burgholzerRandomStimuliGeneration2021,
burgholzerHandlingNonunitariesQuantum2022,
chun-yuAccurateBDDbasedUnitary2022,
taoGiallarPushbuttonVerification2022,
willeVerificationQuantumCircuits2022,
wangXQDDbasedVerificationMethod2008,
hongApproximateEquivalenceChecking2021,
chenAutomatabasedFrameworkVerification2023,
liuRobustApproachDetecting2023}.	

These design tasks demonstrate the wide variety of software tools operating on and across different levels of \mbox{abstraction---leading} to comprehensive quantum circuit design flows realized through toolkits such as IBM's Qiskit~\cite{qiskit}, Google's Cirq~\cite{cirq}, or Rigetti's Forest~\cite{rigetti}.
At the same time, all these tasks have in common that they have an immensely large complexity. 
Therefore, a multitude of techniques have been proposed for each task---each with its own \mbox{trade-off} between resource demand and quality of the result. It is key in the development of scientific methods to empirically evaluate and compare their performances on practical, relevant benchmarks.

\section{Benchmarking}
\label{sec:benchmarking}
The benchmark suite proposed in this paper is not the first and certainly not the last collection of quantum circuit benchmarks. In this section, we review some of the existing suites and their respective foci. Afterwards, we discuss how \mbox{MQT Bench} further complements this.

\subsection{Current State of the Art}\label{sec:sota}
Providing a comprehensive set of benchmarks to satisfy the needs across all levels of the quantum circuit compilation flow is a challenging task. Several approaches to provide benchmark suites for some of the levels within the quantum circuit compilation flow have already been proposed and a \mbox{non-exhaustive} overview is given in the following:

\emph{Application-Oriented Performance Benchmarks for Quantum Computing} \cite{lubinski2021applicationoriented}: Starting with an example located on the algorithmic level, this benchmark suite rather focuses on \mbox{high-level} descriptions. At the moment of writing, it provides 13 benchmark algorithms which can be generated with a variable number of qubits via jupyter notebooks using multiple software compilation flows and are classified into four categories of complexity. The resulting circuits are available as \mbox{high-level} Python objects without any further compilation steps.

\emph{SupermarQ} \cite{supermarq}: Similarly, SupermarQ also focuses on the algorithmic level by providing benchmarks with an adjustable qubit range for eight algorithms as \mbox{high-level} Python objects via a Python package. Additionally, six feature vectors are proposed to describe the characteristics of the benchmarks.

\emph{QASMbench} \cite{li2022qasmbench}: This benchmark suite focuses on the \mbox{target-independent} level. It offers numerous quantum circuits with a wide but fixed range of both the number of qubits and depth. It classifies the benchmarks into three categories of sizes and three categories of algorithm classes.
All circuits are available in an intermediate representation according to the \mbox{openQASM 2.0} specification~\cite{cross_open_2017}. 

\emph{RevLib} \cite{revlib}: This benchmark suite provides a wide variety of reversible circuits in an intermediate representation format specified by the authors. 
Classical functions are frequently embedded into reversible circuits in order to use them in quantum algorithms, e.g., oracle functions or Boolean building blocks such as the modular exponentiation in Shor's algorithm~\cite{shor}.
In addition, decomposed versions of this benchmark suite have found widespread use in evaluating the performance of mapping tools, e.g., in \cite{revlib_example_2, tanOptimalLayoutSynthesis2020, 
liTacklingQubitMapping2019}.
Since reversible circuits merely form a subclass of quantum circuits and do not employ quantum mechanical effects such as superposition and entanglement, they do not serve as adequate benchmarks for the whole stack.

\subsection{Motivation}
All of the mentioned benchmark suites have in common that they each target a specific abstraction level within the quantum circuit compilation flow. Although the respectively intended target level is well covered, the demand for a \mbox{cross-level} benchmark suite is not fully met yet.
Additionally, so far, there is no mutual consolidation which benchmarks to use for empirical evaluations of software tools. 
This results in a lower comparability, reproduceability, and transparency of results and may even cause confusion because benchmarks believed to realize the same particular task are quite frequently realized in a completely different fashion, e.g., a benchmark called "grover" could realize Grover's algorithm with any particular oracle. 
Consequently, the demand for a benchmark suite that spans the whole stack of abstraction levels constitutes the main motivation of this contribution.
In order to realize such a benchmark suite, several challenges have to be tackled:
\begin{itemize}
\item Distinct requirements for benchmark circuits on different levels: The closer a level is to the actual computing hardware, the more requirements have to be fulfilled, e.g., only gates from the device's native \mbox{gate-set} are allowed and the connectivity of the hardware architecture must be considered. 
\item Accessibility: In order to facilitate its adoption, the benchmark suite needs to be as easy to use as possible. While most benchmark suites already provide \mbox{open-source} access, running and adapting the code to the user's needs, e.g., obtaining a selection of relevant benchmarks, frequently is tedious or impossible and a solution that keeps this hurdle low is desirable.
\item Generalizability of a software tool: This can only be guaranteed if it is evaluated on a broad set of benchmarks,
relating to several characteristics, e.g., the number of qubits, circuit depth, or the application domain. 
\item Extendability: In addition to that, quantum computing is rapidly advancing and the extendability of the benchmark suite to future algorithms, \mbox{gate-sets}, and architectures becomes even more important.
\end{itemize}

\mbox{MQT Bench} has been developed with these challenges in mind and aims to be a first step towards covering the whole quantum software stack with a benchmark suite based on the following four core traits (which are described in detail in \autoref{sec:detail_section}):
\begin{enumerate}
\item Cross-level benchmarking: All benchmarks are provided on four abstraction levels reviewed in \autoref{sec:compflow}.
\item Accessibility: A website is provided (\url{https://www.cda.cit.tum.de/mqtbench}) to simplify the usage of \mbox{MQT Bench} as much as possible. 
Additionally, all benchmarks can also be generated \mbox{on-demand} using our Python package.
While most users' needs should be covered by this, we also give access to the \mbox{open-source} repository on GitHub (\url{https://github.com/cda-tum/mqt-bench}).
\item Algorithm Selection: Broad selection of different benchmarks with parameterizable characteristics ranging from building blocks, i.e., QFT, to applications, i.e., Grover's algorithm.
\item Extendability: \mbox{MQT Bench} is easily extendable with respect to available benchmarks, native \mbox{gate-sets}, and hardware architectures.
\end{enumerate}

Benchmarking quantum software tools and compilation flows with the same benchmarks aims to aid comparability, reproducibility, and transparency of empirical evaluations. 
How this leads to more consistent testing is shown in the following example.

\begin{figure}[t]
\begin{center}
\begin{tikzpicture}[node distance=1cm, auto,]
\tikzstyle{every node}=[font=\small]
 \node[punkt] (layer_1) {Algorithmic};
  \node[punkt, inner sep=2pt,below=0.1cm of layer_1] (layer_2) {Target-independent};
   \node[punkt, inner sep=2pt,below=0.1cm of layer_2](layer_3) {Target-dependent native gates};
    \node[punkt, inner sep=2pt,below=0.1cm of layer_3] (layer_4) {Target-dependent mapped};
 \node[rectangle, left=2.5cm of layer_1, draw, align=center, minimum width=3.0cm](comp_1) at ($(layer_1)!0.5!(layer_2)$){Appl.-orient. \\Perf. Benchm. \cite{lubinski2021applicationoriented} \\SupermarQ \cite{supermarq}};
 \node[rectangle, below=0.3cm of comp_1, draw, align=center, minimum width=3.0cm](comp_2){QASMbench \cite{li2022qasmbench}};
 \node[rectangle, below=0.3cm of comp_2, draw, minimum width=3.0cm](comp_3) {RevLib \cite{revlib}};
 \node[rectangle, right=1.7cm of layer_1, draw](MQTbench)  at ($(layer_2)!0.5!(layer_3)$) {MQT Bench};

\draw (MQTbench.west) -- (1.5,-1.47);
\draw[->] (1.5,-1.47) |- (layer_1.east);
\draw[->] (1.5,-1.47) |- (layer_2.east);
\draw[->] (1.5,-1.47) |- (layer_3.east);
\draw[->] (1.5,-1.47) |- (layer_4.east);

\draw[->] (comp_1.east) -- (-1.8, -0.45) |- (layer_1.west);
\draw[->, dashed, transform canvas={yshift=4.0pt}] (-1.8, -0.45) |- (layer_2);
\draw[->, dashed, transform canvas={yshift=4.0pt}]  (-1.8, -0.45) |- (layer_3);
\draw[->, dashed, transform canvas={yshift=4.0pt}] (-1.8, -0.45) |- (layer_4);

\draw (comp_2.east) -- (-2.0, -1.69);
\draw[->, transform canvas={yshift=-4.0pt}] (-2.0, -1.69) |- (layer_2);
\draw[->, dashed, transform canvas={yshift=-0.0pt}] (-2.0, -1.69) |- (layer_3);
\draw[->, dashed] (-2.0, -1.69) |- (layer_4);

\draw (comp_3.east) -- (-2.2, -2.555) ;
\draw[->, transform canvas={yshift=0.0pt}] (-2.2, -2.64) |- (layer_2);
\draw[->, dashed, transform canvas={yshift=-4.0pt}] (-2.2, -2.615) |- (layer_3);
\draw[->, dashed, transform canvas={yshift=-4.0pt}] (-2.2, -2.615) |- (layer_4);

\end{tikzpicture}
\centering

{\solidLegend} Directly applicable\\
{\dashedLegend} Additional effort needed

\caption{Benchmarking software tools with and without \mbox{MQT Bench}.} 
\label{fig:testing_qtum}
\end{center}
\end{figure}
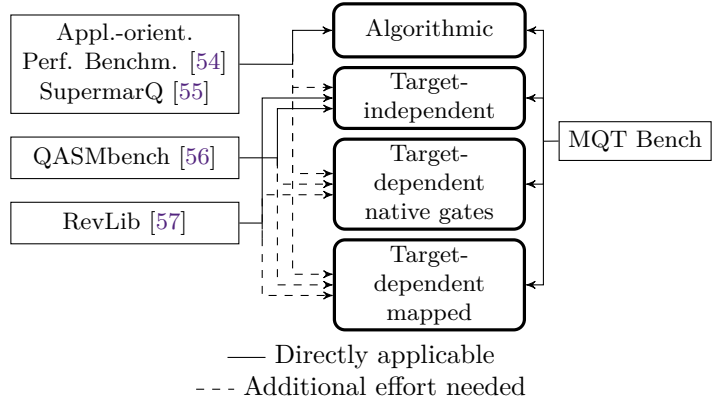

\begin{example}
Consider a full quantum compilation flow across all four levels which shall be developed and tested as depicted in \autoref{fig:testing_qtum}. On the left-hand side of the levels, the current \mbox{state of the art} of benchmarking is sketched. To comprehensively test each level, different benchmark suites have to be adapted and used. In contrast, \mbox{MQT Bench} replaces all of those and enables testing software tools with the same benchmarks throughout the whole quantum circuit compilation flow as shown on the right-hand side.
\label{ex:testing_qtum}
\end{example}

\vspace{5cm}
\section{MQT Bench}
In this section, we describe in more detail how each of the four core traits mentioned above is implemented.
\label{sec:detail_section}
\subsection{Cross-Level Benchmarking}
As exemplarily illustrated in \autoref{sec:background}, software development in quantum computing takes place on various levels. 
While there arguably are numerous possible levels, \mbox{MQT Bench} focuses on four as a balanced measure between specificity and generizability---inspired by the structure proposed in the openQASM~3.0 specification~\cite{cross_openqasm_2021}. 
Providing these benchmarks on all of the four abstraction levels is the main contribution of this work.
More precisely, the following levels are covered:
\begin{enumerate}
\item \emph{Algorithmic level}: In this format, all kinds of quantum gates may be used and subsumed into \mbox{high-level} building blocks. Loops are possible and the circuit structure might be described by constants, variables, and mathematical expressions. This level provides benchmarks on the most generic level and specific values are assigned to its variables, e.g., the number of iterations of a loop. The number of qubits can be specified.
\item \emph{Target-independent level}: Here, loop unrolling, constant folding and propagation are conducted. Also, naive gate simplification is enforced. Again, the number of qubits can be adapted through the website and the used compiler can be selected.
\item \emph{Target-dependent native gates level}: On this level, a particular native \mbox{gate-set} is chosen and the circuit is transpiled and optimized accordingly.
Here, the native \mbox{gate-set}, the compiler used and its settings can be specified.
\item \emph{Target-dependent mapped level}: Finally, a dedicated architecture layout is chosen and the circuit is mapped to the targeted device such that it satisfies all its connectivity constraints and becomes executable on the device. 
Similar to the previous level, the used compiler and its settings are selectable---in addition to the target device.

\end{enumerate}

All relevant information of a generated benchmark is denoted in short in the file name and in detail within the file itself.
\begin{figure*}[t]
\centering
\includegraphics[width=1.0\linewidth]{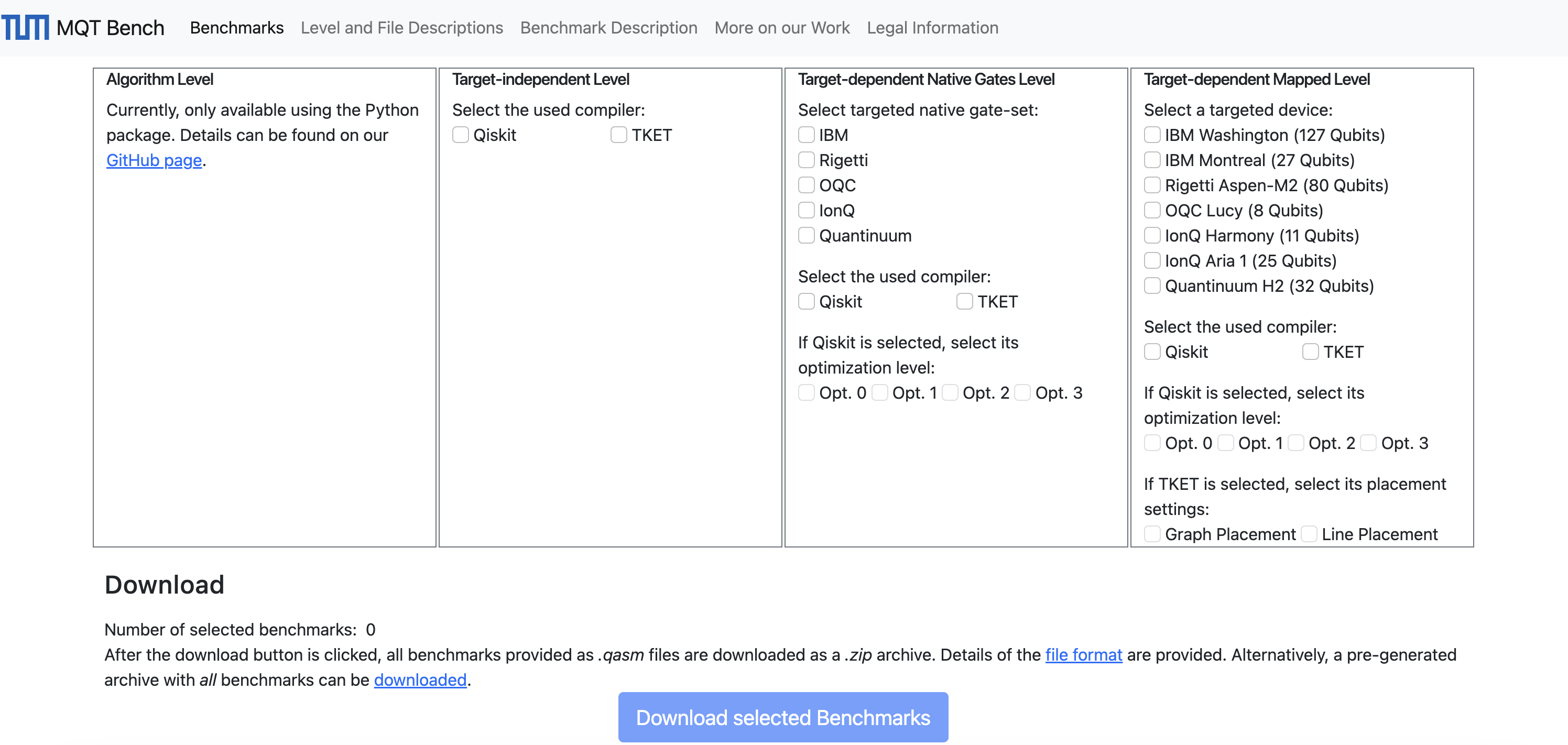}
\caption{User interface of the MQT Bench website.}
\label{fig:website_1}
\end{figure*}

\vspace{5cm}
\subsection{Accessibility}
\mbox{MQT Bench} strives to meet the needs of many. 
To accomplish this goal, it has to be as \mbox{user-friendly} as possible. 
On the one hand, this involves the benchmark library itself. 
All of the \mbox{high-level}, algorithmic benchmark scripts and routines to generate representations for the individual levels are publicly available on GitHub (\url{https://github.com/cda-tum/mqt-bench}).
This allows for complete transparency on how the respective circuits have been generated.
Additionally, a Python package is provided such that each benchmark can be generated \mbox{on-demand}.

On the other hand, and most importantly, this involves the way in which users interface with the benchmark suite. 
While most benchmark suites are provided as an accumulation of benchmark files---making it hard to extract the desired set of benchmarks, \mbox{MQT Bench} includes an \mbox{easy-to-use}, \mbox{no-coding-required} web interface (\url{https://www.cda.cit.tum.de/mqtbench}) that provides the means to filter the large number of pre-generated benchmarks according to the specific needs of the user. 
This web interface and its underlying server software is also part of our Python package, such that every user can utilize the same interface locally without the need to access our publicly available server.
A screenshot of the website's user interface and its configuration options is shown in \autoref{fig:website_1}.

\subsection{Algorithm Selection}
To provide a broad spectrum of different benchmarks, \mbox{MQT Bench} comprises most of today's \mbox{de-facto} standard quantum algorithms. 
This includes building blocks, e.g., Quantum Fourier Transform (QFT) and the Greenberger-Horne-Zeilinger state preparation (GHZ), up to \mbox{higher-level} algorithms, e.g., Grover's~\cite{grover} and Shor's~\cite{shor} algorithm.
At the time of writing, \mbox{MQT Bench} comprises the following benchmarks:
\begin{itemize}
\item Amplitude Estimation (AE)
\item Deutsch-Jozsa (DJ) algorithm
\item Graph state preparation
\item GHZ state preparation
\item Grover's algorithm
\item Quantum Approximation Optimization Algorithm (QAOA)
\item Quantum Fourier Transformation (QFT)
\item Quantum Phase Estimation (QPE)
\item Quantum Walk
\item Random circuit
\item Shor's algorithm
\item Variational Quantum Eigensolver (VQE)
\item Three VQA ansatz functions with randomly initialized parameters: Two Local, Real Amplitudes and Efficient SU2
\item W-State preparation
\end{itemize}

Additionally, MQT Bench provides several application benchmarks specifically targeting variational quantum algorithms, since those algorithms are especially promising in the \mbox{NISQ-era \cite{cerezo_variational_2021}}. To this end, we use the classification provided by IBM Qiskit's application levels: Optimization, Machine Learning, Finance, and Nature:
\begin{itemize}
\item \emph{Optimization}: Travelling salesman problem and vehicle routing
\item \emph{Machine Learning}: Quantum Neural Network (QNN)
\item \emph{Finance}: Portfolio optimization and option pricing
\item \emph{Nature}: Ground state estimation
\end{itemize}

\subsection{Extendability}
Quantum computing is a rapidly evolving area of research and, especially in recent years, numerous promising algorithms and applications have emerged. 
Any benchmark suite has to be extendable and must continuously evolve in order to remain relevant. \mbox{MQT Bench} is designed to be easily extendable in a multitude of ways:

\begin{itemize}
\item \emph{Algorithms}: New applications and algorithms can be easily integrated into \mbox{MQT Bench} by providing the corresponding algorithmic level description. The generation of all other levels and options (such as the native \mbox{gate-sets}, architectures, and compilers with their settings) is automatically handled by the proposed library.
\item \emph{Native gate-sets}: So far, native \mbox{gate-sets} of superconducting quantum computers from IBM, Rigetti, and Oxford Quantum Circuits are provided---in addition to the \mbox{gate-sets} of IonQ's and Quantinuum's ion \mbox{trap-based} quantum computers.
In the future, additional \mbox{gate-sets} (such as those of Google's and or AQT's quantum computers) can be realized by adopting the necessary compilation routines in the proposed library.
\item \emph{Hardware architectures}: All major players currently rely on a modular platform for the architecture of their devices in order to further scale their quantum computers. It is straightforward to extend \mbox{MQT Bench's} list of available architectures to accommodate future architectures.
\end{itemize}

Although the four levels \mbox{MQT Bench} is based on cover today's main use cases, the number of abstraction levels most likely is going to increase in the future. While the mapped level has been considered the last one before a circuit is sent to the quantum computer for execution, \mbox{pulse-level} programming is envisioned to give even more control to software developers as proposed in \cite{Li2022pulselevelnoisy}. On the other end of the spectrum, \mbox{higher-level} abstractions and programming languages will be required to foster the adoption of quantum technology. Currently, a programmatic description of the benchmarks is needed on the algorithmic level. An even higher level where this programmatic description will be automatically derived from is also envisioned. 
\emph{QUARK}~\cite{quark}, a framework for quantum computing application benchmarking consisting of four levels where the lowest level considers the whole compilation flow discussed in \autoref{sec:compflow}, constitutes one step towards this direction.   

\begin{figure*}[t]
\centering
\begin{subfigure}{0.5\textwidth}
    \includegraphics[width=\textwidth]{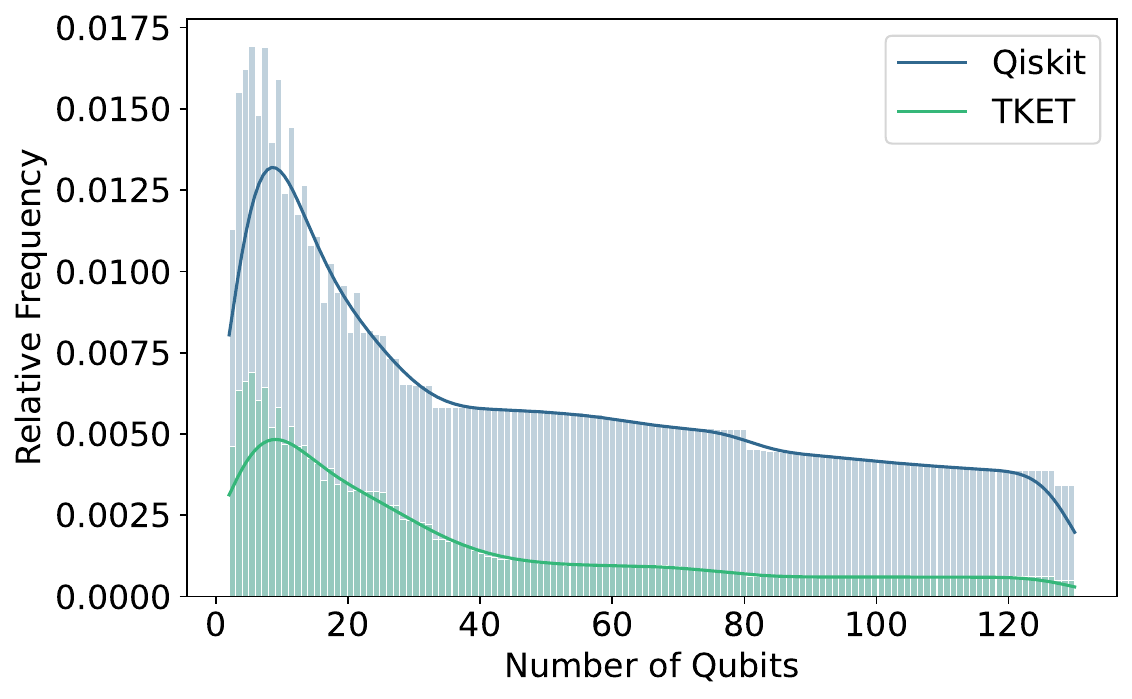}
    \caption{Relative distribution of benchmarks per number of qubits and used compiler.}
    \label{fig:dist_num_qubits}
\end{subfigure}
\hfill
\begin{subfigure}{0.49\textwidth}
    \includegraphics[width=\textwidth]{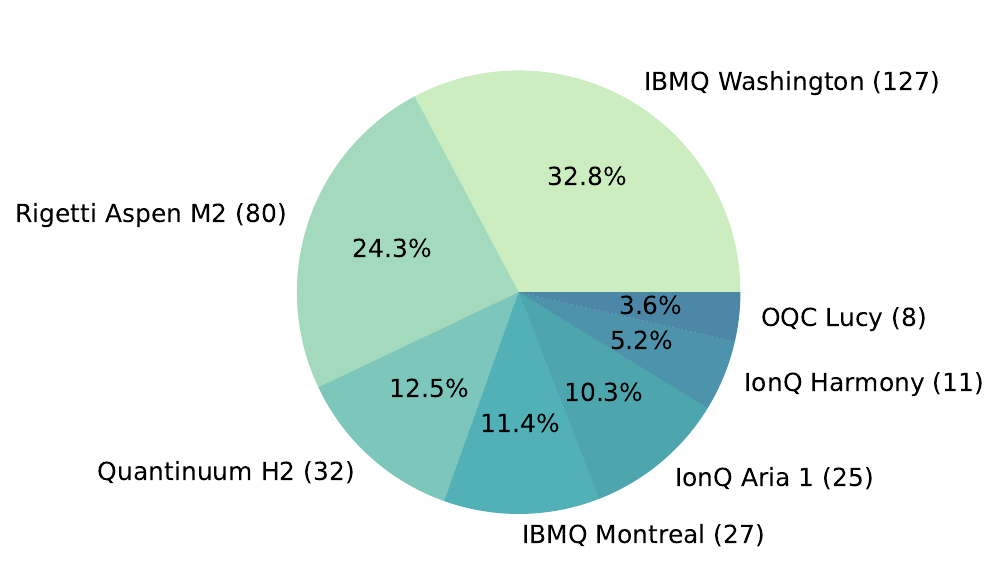}
    \vfill
    \caption{Device distribution of target-dependent mapped level benchmarks.}
    \label{fig:dist_devices}
\end{subfigure}
\hfill
\begin{subfigure}{0.99\textwidth}
    \includegraphics[width=\textwidth]{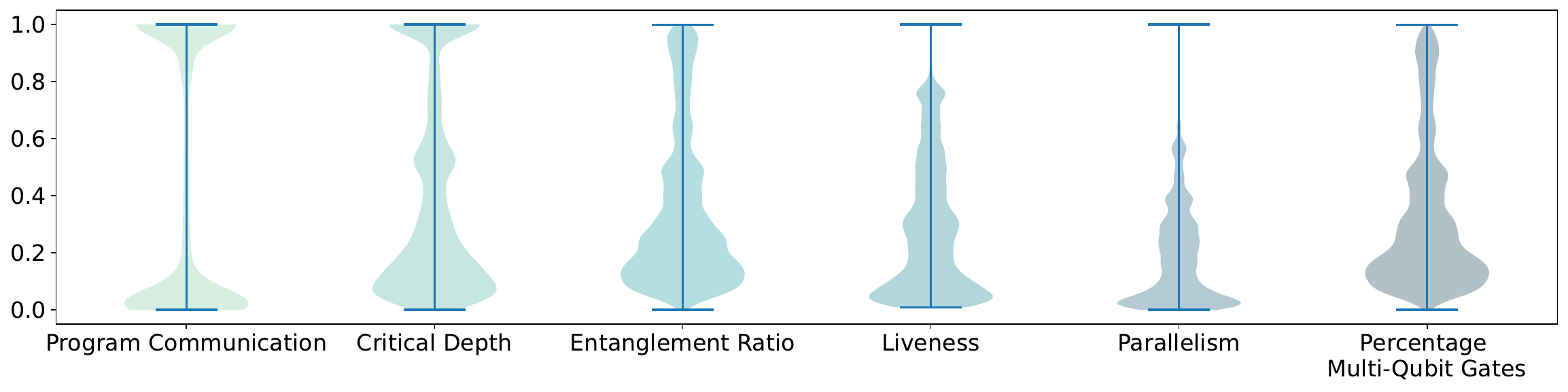}
    \caption{Distributions of quantum circuit characteristics (with all but the last one taken from~\cite{supermarq}).}
    \label{fig:dist_violins}
\end{subfigure}
\caption{Insights into the provided benchmarks and their characteristics.}
\label{fig:evaluations}
\vspace{-2mm}
\end{figure*}

\section{Evaluation}
\label{sec:eval}
At the time of writing, MQT Bench (\emph{v1.0.0}) considers two compilers, five native \mbox{gate-sets}, and seven devices ranging from $8$ to $127$ qubits. 
In this section, we provide an overview of several characteristics and statistics of the resulting \mbox{pre-generated} benchmarks.
To this end, the following metrics are illustrated in~\autoref{fig:evaluations}:

\emph{Number of qubits}: A first broad overview is given in \autoref{fig:dist_num_qubits}, which shows the relative frequency of the generated benchmarks per number of qubits and per compiler.
The relative frequency of benchmarks is decreasing with an increasing number of qubits due to fewer devices being available for higher qubit numbers and the maximal generation time for a benchmark file being exceeded more frequently. 
Furthermore, the benchmark generation using the \emph{TKET} compiler generally took more time than \emph{Qiskit}---leading to a smaller number of generated benchmarks, especially on the target-dependent mapped level.

\emph{Distribution of target-dependent mapped level benchmarks}: The distribution of the target device for the mapped level is shown in \autoref{fig:dist_devices}, with the number of qubits denoted in brackets.
As expected, the larger the device, the more benchmarks are created for it.
On the target-dependent native gates level, there is an equal distribution with respect to the number of benchmarks created for each of the five native \mbox{gate-sets}.

\emph{Distribution of benchmark characteristics}: In \autoref{fig:dist_violins}, six different characteristics with values between $0$ and $1$ are evaluated for all \mbox{pre-generated} benchmarks.
Five of those characteristics (Program Communication, Critical Depth, Entanglement Ratio, Liveness, and Parallelism) have been proposed in~\cite{supermarq}.
Furthermore, the percentage of \mbox{multi-qubit} gates is evaluated.

\section{Conclusion}\label{sec:conclusion}
In this work, we proposed \emph{MQT Bench}---a quantum circuit benchmark suite (as part of the \emph{Munich Quantum Toolkit}, MQT) comprising different algorithms, compilers, native \mbox{gate-sets}, and target devices---resulting in more than $70{\small,}000$ benchmark circuits ranging from $2$ to $130$ qubits on four abstraction levels.
To keep this large number of benchmarks manageable, we provide an \mbox{easy-to-use} web interface (\url{https://www.cda.cit.tum.de/mqtbench}) allowing users to filter the benchmarks according to their needs. 
\emph{MQT Bench} is also provided as a Python package (including the server software to start the web interface locally), such that each of the benchmarks can be easily generated \mbox{on-demand}.
Furthermore, we give access to the \mbox{open-source} repository on GitHub (\url{https://github.com/cda-tum/mqt-bench}).
By this, \emph{MQT Bench} presents a first step towards serving a single benchmark suite for the whole quantum software stack---facilitating empirical evaluations of quantum software tools that are comparable, reproducible, and transparent.

\section*{Acknowledgments}
This work received funding from the European Research Council (ERC) under the European Union’s Horizon 2020 research and innovation program (grant agreement No. 101001318), was part of the Munich Quantum Valley, which is supported by the Bavarian state government with funds from the Hightech Agenda Bayern Plus, and has been supported by the BMWK on the basis of a decision by the German Bundestag through project QuaST, as well as by the BMK, BMDW, and the State of Upper Austria in the frame of the COMET program (managed by the FFG).

\newpage
\section*{References}
\sloppy

\printbibliography[heading=none]

\end{document}